\documentclass[a4paper,fleqn,usenatbib]{mnras}

\usepackage{newtxtext,newtxmath}

\usepackage[T1]{fontenc}
\usepackage{ae,aecompl}

\usepackage{amsmath}
\usepackage{graphicx}
\usepackage{epstopdf}
\usepackage{latexsym}
\usepackage{amssymb}
\usepackage{hyperref}
\usepackage{comment}
\hypersetup{
    colorlinks=true,
    linkcolor=blue,
    citecolor=blue,
    urlcolor=blue
}
\usepackage{array}
\usepackage{enumitem}
\usepackage{float}

\title[An interacting dark sector and GW170817]{An interacting dark sector and the first gravitational-wave standard siren detection}

\author[J. Mifsud and C. van de Bruck]{
Jurgen Mifsud\thanks{E-mail: jmifsud1@sheffield.ac.uk}
and Carsten van de Bruck
\\
Consortium for Fundamental Physics, School of Mathematics and Statistics, University of Sheffield, Hounsfield Road, Sheffield\\ S3 7RH, UK
}

\date{Accepted XXX. Received YYY; in original form ZZZ}

\pubyear{2019}

\begin{document}
\label{firstpage}
\pagerange{\pageref{firstpage}--\pageref{lastpage}}
\maketitle

\begin{abstract}
After the first nearly simultaneous joint observations of gravitational--waves and electromagnetic emission produced by the coalescence of a binary neutron star system, another probe of the cosmic expansion which is independent from the cosmic distance ladder, became available. We perform a global analysis in order to constrain an interacting dark energy model, characterised by a conformal interaction between dark matter and dark energy, by combining current data from: \textit{Planck} observations of the cosmic microwave background radiation anisotropies, and a compilation of Hubble parameter measurements estimated from the cosmic chronometers approach as well as from baryon acoustic oscillations measurements. Moreover, we consider two measurements of the expansion rate of the Universe today, one from the observations of the Cepheid variables, and another from the merger of the binary neutron star system GW170817. We find that in this interacting dark energy model, the influence of the local measurement of the Hubble constant mostly affects the inferred constraints on the coupling strength parameter between dark energy and dark matter. However, the GW170817 Hubble constant measurement is found to be more conservative than the Cepheid variables measurement, and in a better agreement with the current high redshift cosmological data sets. Thus, forthcoming gravitational--wave standard siren measurements of the Hubble constant would be paramount for our understanding of the dark cosmic sector.
\end{abstract}




\section{Introduction}
\label{sec:introduction}
Gravitational--wave multi--messenger astronomy paved the way for the possibility of using standard sirens to infer the current expansion rate of our Universe. It has long been acknowledged (see, for instance, \citet{Schutz:1986gp,Krolak:1987ofj,Chernoff:1993th,Markovic:1993cr,Finn:1995ah,Wang:1996in,Thorne:1997ut,Zhu:2001mb,Holz:2005df,Dalal:2006qt,Taylor:2011fs,2013arXiv1307.2638N}) that gravitational--wave inspiral detections would provide us with invaluable cosmological information. Since the amplitude of a binary's gravitational--wave signal encodes its luminosity distance \citep{Congedo:2017qbq}, binary inspirals became known as standard sirens \citep{Schutz:1986gp}, which are the gravitational--wave analogues of type Ia supernovae standard candle measurements. In particular, the determination of the Hubble constant from gravitational--wave standard sirens \citep{Schutz:1986gp,Krolak:1987ofj,Chernoff:1993th,Finn:1995ah,Dalal:2006qt,Taylor:2011fs,2013arXiv1307.2638N} was demonstrated for the first time by the nearly concurrent joint observations of the electromagnetic counterpart (see \citet{GBM:2017lvd,Monitor:2017mdv,Goldstein:2017mmi,Soares-Santos:2017lru,Coulter:2017wya,Savchenko:2017ffs,Valenti:2017ngx,Arcavi:2017xiz,Tanvir:2017pws}, and references therein) to the gravitational--wave signal \citep{TheLIGOScientific:2017qsa} produced by the merger of the binary neutron star system GW170817 which has been localised to the host galaxy NGC 4993.
\\[2pt]\indent
Although the first constraint on the Hubble constant from standard sirens \citep{Abbott:2017xzu} is significantly weaker than the inferred constraints from observations of Cepheid variables (see \citet{Riess:2016jrr}, and the new analysis of \citet{2018ApJ...855..136R,Riess:2018byc}) and the extrapolated concordance model cosmic microwave background (CMB) measurement \citep{Aghanim:2016yuo} (see also \citet{Ade:2013zuv,Ade:2015xua,Aghanim:2018eyx}), prospective gravitational--wave standard siren measurements of the Hubble constant are expected to be significantly improved after the detection of additional standard siren events. Consequently, these near future standard siren measurements of the Hubble constant would be competitive with the measurements inferred from the more established methods \citep{Chen:2017rfc,Hotokezaka:2018dfi,Feeney:2018mkj}. Moreover, standard siren measurements of the Hubble constant are independent of the cosmic distance ladder or poorly understood calibration processes, as these are primarily calibrated by the robust theory of General Relativity to cosmological scales and instrumental systematics are expected to be inconsequential \citep{Karki:2016pht,Cahillane:2017vkb,Davis:2018yrz}. Furthermore, we should also point out that the reported standard siren constraint of $H_0=70^{+12}_{-8}$ km s$^{-1}$ Mpc$^{-1}$ at the 68\% confidence level is strongly non--Gaussian \citep{Abbott:2017xzu}, with the major uncertainty being the inclination plane of the binary orbit. This independent probe of the present--day cosmic expansion is paramount for the reported discrepancy at the ($\ga$)3$\sigma$ level \citep{Feeney:2017sgx} between the locally measured \citep{Riess:2016jrr} and the CMB derived estimate \citep{Aghanim:2016yuo} of the Hubble constant, as forthcoming standard siren detections would be able to adjudicate between these discrepant measurements \citep{Feeney:2018mkj}. Such disagreement could either be an indication of several physical mechanisms beyond our concordance model of cosmology (see, for instance, \citet{Odderskov:2015fba,DiValentino:2016hlg,DiValentino:2017iww,DiValentino:2017rcr,Qing-Guo:2016ykt,Grandis:2016fwl,Karwal:2016vyq,Bernal:2016gxb,Lancaster:2017ksf,Prilepina:2016rlq,Zhao:2017urm,Sola:2017znb,Colgain:2018wgk,vandeBruck:2017idm,Poulin:2018zxs}), or unidentified systematic errors (see \citet{Addison:2015wyg,Cardona:2016ems,Zhang:2017aqn,Odderskov:2017ivg,Wu:2017fpr,Feeney:2017sgx,Follin:2017ljs,Dhawan:2017ywl,Camarena:2018nbr}, and references therein), although there is still no compelling explanation to date.

Given that the derived Hubble constant measurement from the CMB assumes a $\Lambda$CDM cosmic evolution, in which the cosmological expansion is dominated by a cosmological constant ($\Lambda$) and cold dark matter (CDM), a number of alternative cosmological models have been proposed. For instance, models with a time--evolving \citep{Zhao:2017cud,DiValentino:2017rcr} along with other non--standard dark energy cosmic components \citep{Karwal:2016vyq,Qing-Guo:2016ykt,DiValentino:2017zyq,DiValentino:2017iww,Yang:2018euj,Yang:2018ubt}, and neutrino contributions \citep{Riess:2016jrr,Kumar:2016zpg,Ko:2016uft,Archidiacono:2016kkh,Yang:2017amu,Zhao:2017urm,DiValentino:2017oaw,Benetti:2017juy} have been shown to partially alleviate this Hubble constant tension reported in the $\Lambda$CDM framework. Thus, independent gravitational--wave standard siren measurements of the Hubble constant would certainly shed light on the physics beyond the concordance cosmological model, particularly when the sub--percent level is attained. Such accurate standard siren measurements were repeatedly shown that these will be able to constrain the cosmological parameters (see, for instance, \citet{Dalal:2006qt,MacLeod:2007jd,Cutler:2009qv,Sathyaprakash:2009xt,Zhao:2010sz,Nishizawa:2011eq,DelPozzo:2011yh,Taylor:2012db,Tamanini:2016zlh,Belgacem:2018lbp,DiValentino:2018jbh,Feeney:2018mkj,Congedo:2018wfn}), and would be of utmost importance for the forthcoming CMB and baryon acoustic oscillation (BAO) surveys which are expected to reach an unprecedented level of accuracy \citep{Abazajian:2016yjj,DiValentino:2016foa}.

We should also remark that apart from the Hubble constant measurement, the observations of gravitational--wave and electromagnetic emission from the coalescence of the binary neutron star system GW170817 have been used to test our understanding of gravitation and astrophysics \citep{Lombriser:2015sxa,Monitor:2017mdv,Abbott:2018lct}. For instance, the fractional speed difference between the speed of light and that of gravity has been exquisitely found to be less than about one part in 10$^{15}$ \citep{Monitor:2017mdv}, which consequently led to stringent constraints on several modified theories of gravity (see, for instance, \citet{Baker:2017hug,Creminelli:2017sry,Sakstein:2017xjx,Ezquiaga:2017ekz,Langlois:2017dyl,Dima:2017pwp,deRham:2018red}).

It is therefore timely to investigate the impact of the first gravitational--wave standard siren measurement of the Hubble constant on the current CMB and cosmic expansion constraints in the framework of a cosmological model characterised by a non--standard interacting dark sector. A similar analysis has been carried out in an extended $\Lambda$CDM model \citep{DiValentino:2017clw}, in which the inclusion of the GW170817 Hubble constant measurement led to improved constraints on the model parameters. We here consider a cosmological model in which dark matter and dark energy interact with one another, whereas the standard model particles follow their standard cosmological evolution. Consequently, this coupled dark energy model evades the tight constraints inferred from the equivalence principle and solar system tests \citep{Bertotti:2003rm,Will}. Due to the obscure nature of dark matter and dark energy, a dark sector coupling cannot be excluded from the viewpoint of fundamental physics \citep{Damour:1990tw,Wetterich:1994bg,Carroll:1998zi,Holden:1999hm,Gubser:2004du,Farrar:2003uw,Carroll:2008ub}, and such an interaction between these dark sector constituents is not currently forbidden by cosmological data (see, for instance, \citet{Salvatelli:2014zta,Kumar:2016zpg,Kumar:2017dnp,Abdalla:2014cla,vandeBruck:2016hpz,vandeBruck:2017idm,Yang:2018ubt}). We here consider an interacting dark energy model in which an evolving dark energy scalar field \citep{Wetterich:1987fm,Ratra:1987rm,Peebles:1987ek} is coupled to the dark matter quanta via the so--called conformal coupling function, and is characterised by a dark sector fifth--force between the dark matter particles mediated by the dark energy scalar field. The modified cosmological evolution along with its distinct cosmological signatures on the linear and non--linear levels have been exhaustively explored in the literature (see, for instance, \citet{Wetterich:1994bg,Amendola:1999er,Amendola:2003wa,Farrar:2003uw,Mainini:2006zj,Pettorino:2008ez,Baldi:2008ay,Baldi:2010vv,Baldi:2010pq,Baldi:2011th,Baldi:2011qi,Jack,Odderskov:2015fba,Mifsud:2017fsy}), and tight constraints on the model parameters have been placed \citep{Amendola:2003eq,Bean:2008ac,Xia:2009zzb,Amendola:2011ie,Pettorino:2012ts,Pettorino:2013oxa,Xia:2013nua,Ade:2015rim,Miranda:2017rdk,vandeBruck:2017idm}. Thus, the aim of our analysis is to compare the impact of the Hubble constant measurement derived from the binary neutron star system GW170817 with that of the locally inferred Hubble constant measurement on these tight model parameter constraints.

The organisation of this paper is as follows. In Section \ref{sec:model} we briefly introduce the considered interacting dark energy model, and in Section \ref{sec:Data Sets} we summarise the observational data sets together with the method that will be employed to infer the cosmological parameter constraints. We then present and discuss our results in Section \ref{sec:results}, and draw our final remarks and prospective lines of research in Section \ref{sec:conclusions}.
\section{Interacting Dark Energy}
\label{sec:model}
We here briefly review the basic equations of our interacting dark energy (DE) model. The phenomenology of this dark sector interaction can be immediately grasped by writing down the Einstein frame scalar--tensor theory action:
\begin{equation}\label{action}
\begin{split}
\mathcal{S} =& \int \mathrm{d}^4 x \sqrt{-g} \left[ \frac{M_{\rm Pl}^2}{2} R - \frac{1}{2} g^{\mu\nu}\partial_\mu^{} \phi\, \partial_\nu^{} \phi - V(\phi) + \mathcal{L}_\text{SM}^{}\right]\\
&+ \int \mathrm{d}^4 x \sqrt{-\tilde{g}} \mathcal{\widetilde{L}}_\text{DM}^{}\left(\tilde g_{\mu\nu}^{}, \psi\right),
\end{split}
\end{equation}
in which the gravitational sector has the standard Einstein--Hilbert form, and define $M_\text{Pl}^{-2}\equiv 8\pi G$ such that $M_\text{Pl}^{}=2.4\times 10^{18}$ GeV is the reduced Planck mass. DE is promoted to a dynamical scalar field, as in the vast majority of alternative DE models, and is described by a canonical quintessence scalar field $\phi$, with a potential $V(\phi)$. The uncoupled standard model (SM) particles are depicted by the Lagrangian $\mathcal{L}_\text{SM}^{}$, which incorporates a relativistic and a baryonic sector (hereafter denoted by the subscripts $r$ and $b$, respectively). Particle quanta of the dark matter (DM) fields $\psi$, follow the geodesics defined by the metric $\tilde g_{\mu\nu}^{} = C(\phi) g_{\mu\nu}^{}$, with $C(\phi)$ being the dark sector conformal coupling function\footnote{This metric transformation can be considered as a particular case of a generalised transformation which takes into account a conformal as well as a non--vanishing disformal \citep{Bekenstein:1992pj} dark sector coupling function \citep{Zumalacarregui:2012us,Koivisto,Jack,vandeBruck:2016hpz,Mifsud:2017fsy,vandeBruck:2017idm,Xiao:2018jyl}.}.

As a consequence of the interaction between the dark sector constituents, the modified conservation equations of the energy--momentum tensors of the scalar field and DM are respectively given by
\begin{equation}\label{eq:modKG}
\Box\phi=V_{,\phi} - Q\;,\;\;\;\nabla^\mu_{} T^\text{DM}_{\mu\nu}=Q\nabla_\nu^{}\phi\;,
\end{equation}
where $V_{,\phi}\equiv \mathrm{d}V/\mathrm{d}\phi$. Moreover, the dark sector coupling function is given by
\begin{equation}\label{coupling}
Q=\frac{C_{,\phi}}{2C}T_\text{DM}^{}\;,
\end{equation}
with $T_\text{DM}^{}$ being the trace of the perfect fluid energy--momentum tensor of pressureless DM, denoted by $T^\text{DM}_{\mu\nu}$. As illustrated in equation (\ref{action}), SM particles are excluded from the dark sector interaction, thus their perfect fluid energy--momentum tensor satisfies $\nabla^\mu T^\text{SM}_{\mu\nu}=0$.

On assuming a spatially--flat Friedmann--Lema\^{i}tre--Robertson--Walker (FLRW) line element, specified by $\mathrm{d}s^2 = g_{\mu\nu}\mathrm{d}x^{\mu} \mathrm{d}x^{\nu} = a^2(\tau)\left[-\mathrm{d}\tau^2 + \delta_{ij} \mathrm{d}x^i \mathrm{d}x^j\right]$, the evolution of the DE scalar field is governed by
\begin{equation}\label{KG-equation}
\phi^{\prime\prime} + 2 \mathcal{H} \phi^{\prime} + a^2 V_{,\phi} = a^2 Q\;,
\end{equation}
where a prime denotes a derivative with respect to conformal time $\tau$, and define the conformal Hubble parameter by $\mathcal{H}=a^\prime/a$, with $a(\tau)$ being the cosmological scale factor. Furthermore, the DM energy density $\rho_c^{}$, satisfies an energy exchange equation, given by
\begin{equation}\label{conservation_matter}
\rho_c^\prime + 3\mathcal{H}\rho_c^{} = -Q\phi^{\prime}_{}\;,
\end{equation}
where the coupling function in FLRW simplifies \citep{Wetterich:1994bg,Amendola:1999er,Zumalacarregui:2012us,Jack,Mifsud:2017fsy} to $Q=-C_{,\phi}\rho_c^{}/(2C)$.  Throughout this paper we adopt the following exponential conformal coupling and scalar field potential functions
\begin{equation}\label{coupling_choice}
C(\phi)=e^{2\alpha\phi/M_\text{Pl}}\;,\;\;\;\;V(\phi)=V_0^4 e^{-\lambda\phi/M_\text{Pl}}\;,
\end{equation}
where $\alpha$, $V_0$ and $\lambda$ are constants.
%

Due to the non--negligible cosmological imprints on the evolution of cosmic perturbations and background dynamics, such an interaction within the dark sector has been widely studied and tight constraints were inferred from several cosmological probes (see, for instance, \citet{Amendola:2003eq,Bean:2008ac,Xia:2009zzb,Amendola:2011ie,Pettorino:2012ts,Pettorino:2013oxa,Xia:2013nua,Ade:2015rim,Miranda:2017rdk,vandeBruck:2017idm}, and references therein). We here illustrate the distinctive imprints of two independent Hubble constant measurements on the \textit{Planck} and cosmic expansion constraints, particularly on the allowed conformal coupling strength parameter values.

{\setlength\extrarowheight{5pt}
\setlength{\tabcolsep}{8.5pt}
\begin{table}
\begin{center}
\caption{\label{table:priors} External flat priors on the cosmological parameters assumed in this paper.}
\begin{tabular}{ p{1.85cm}  l } 
 \hline
\hline
Parameter~  & ~Prior~ \\[2.6pt]
\hline
$\Omega_b h^2$\dotfill & $[0.005,\,0.100]$ \\
$\Omega_c h^2$\dotfill & $[0.01,\,0.99]$ \\
100 $\theta_s$\dotfill & $[0.5,\,10.0]$ \\
$\tau_{\mathrm{reio}}$\dotfill & $[0.02,\,0.80]$ \\
$\ln(10^{10}A_s)$\dotfill & $[2.7,\,4.0]$ \\
$n_s$\dotfill & $[0.5,\,1.5]$ \\ 
$\lambda$\dotfill & $[0.0,\,1.7]$ \\ 
$\alpha$\dotfill & $[0.00,\,0.48]$ \\[2pt]
\hline
\hline
\end{tabular}
\end{center}
\end{table}}
{\setlength\extrarowheight{5pt}
\begin{table*}
\begin{center}
\caption{\label{table:Tab1} For each model parameter we report the mean values and $1\sigma$ errors, together with the $1\sigma$ ($2\sigma$) upper limits of $\lambda$ and $\alpha$. The Hubble constant is given in units of km s$^{-1}$ Mpc$^{-1}$.}
\begin{tabular}{ p{1.85cm} c  c  c }
 \hline
\hline
Parameter~  &  ~\textit{Planck}~ & ~$+\,H_0^{\mathrm{GW}}$~ & ~$+\,H_0^{\mathrm{R}}$~  \\[2.5pt]
\hline
100 $\Omega_b h^2$\dotfill & $2.2261^{+0.0166}_{-0.0172}$ & $2.2263^{+0.0163}_{-0.0168}$ & $2.2274^{+0.0168}_{-0.0171}$ \\
$\Omega_c h^2$\dotfill & $0.11747^{+0.00335}_{-0.00185}$ & $0.11735^{+0.00342}_{-0.00175}$ & $0.11356^{+0.00240}_{-0.00245}$ \\
100 $\theta_s$\dotfill & $1.04180^{+0.00032}_{-0.00033}$ & $1.04180^{+0.00032}_{-0.00032}$ & $1.04190^{+0.00032}_{-0.00032}$ \\
$\tau_{\mathrm{reio}}$\dotfill & $0.063591^{+0.014217}_{-0.014531}$ & $0.063398^{+0.014010}_{-0.014284}$ & $0.066223^{+0.013861}_{-0.013988}$ \\
$\ln(10^{10}A_s)$\dotfill & $3.0600^{+0.0267}_{-0.0263}$ & $3.0596^{+0.0258}_{-0.0263}$ & $3.0650^{+0.0254}_{-0.0262}$ \\
$n_s$\dotfill & $0.96651^{+0.00506}_{-0.00539}$ & $0.96658^{+0.00502}_{-0.00529}$ & $0.96906^{+0.00481}_{-0.00512}$ \\ 
$\lambda$\dotfill & $0.81539^{+0.29079}_{-0.81538}$ & $0.77378^{+0.26326}_{-0.77376}$ & $0.52287^{+0.15566}_{-0.52287}$ \\ 
$\alpha$\dotfill & $0.040756^{+0.012186}_{-0.040753}$ & $0.041569^{+0.012589}_{-0.041569}$ & $0.071762^{+0.020766}_{-0.016941}$ \\
$\lambda$\dotfill & $<1.1062(1.6001)$ & $<1.0370(1.5686)$ & $<0.6785(1.1979)$ \\ 
$\alpha$\dotfill & $<0.0529(0.0912)$ & $<0.0542(0.0921)$ & $<0.0925(0.1132)$ \\[2pt]
\hline
$H_0$\dotfill & $67.898^{+2.8019}_{-3.2190}$ & $68.177^{+2.7017}_{-3.0000}$ & $72.040^{+1.8185}_{-1.8632}$ \\
$\Omega_m$\dotfill & $0.30531^{+0.03209}_{-0.03156}$ & $0.30239^{+0.03143}_{-0.02837}$ & $0.26236^{+0.01667}_{-0.01874}$ \\
$\sigma_8^{}$\dotfill & $0.83009^{+0.02638}_{-0.03603}$ & $0.83261^{+0.02471}_{-0.03684}$ & $0.87375^{+0.02403}_{-0.02464}$ \\
$z_{\mathrm{reio}}$\dotfill & $8.5379^{+1.4241}_{-1.2878}$ & $8.5180^{+1.4136}_{-1.2621}$ & $8.7375^{+1.3768}_{-1.2093}$ \\[3.5pt]
\hline
\hline
\end{tabular}
\end{center}
\end{table*}}
%

\section{Data Sets \& Method}
\label{sec:Data Sets}
We now discuss the data sets which are used to confront the above interacting DE model. In all data set combinations we consider the low multipole ($2\leq\ell\leq29$) publicly available \textit{Planck} 2015 data \citep{Aghanim:2015xee}, along with the high multipole ($\ell\geq30$) range, and the \textit{Planck} lensing likelihood in the multipole range $40\leq\ell\leq400$ \citep{Ade:2015zua}. In the following we refer to this combination of CMB angular power spectra as `\textit{Planck}'.

\begin{figure}
\centering
  \includegraphics[width=0.99\columnwidth]{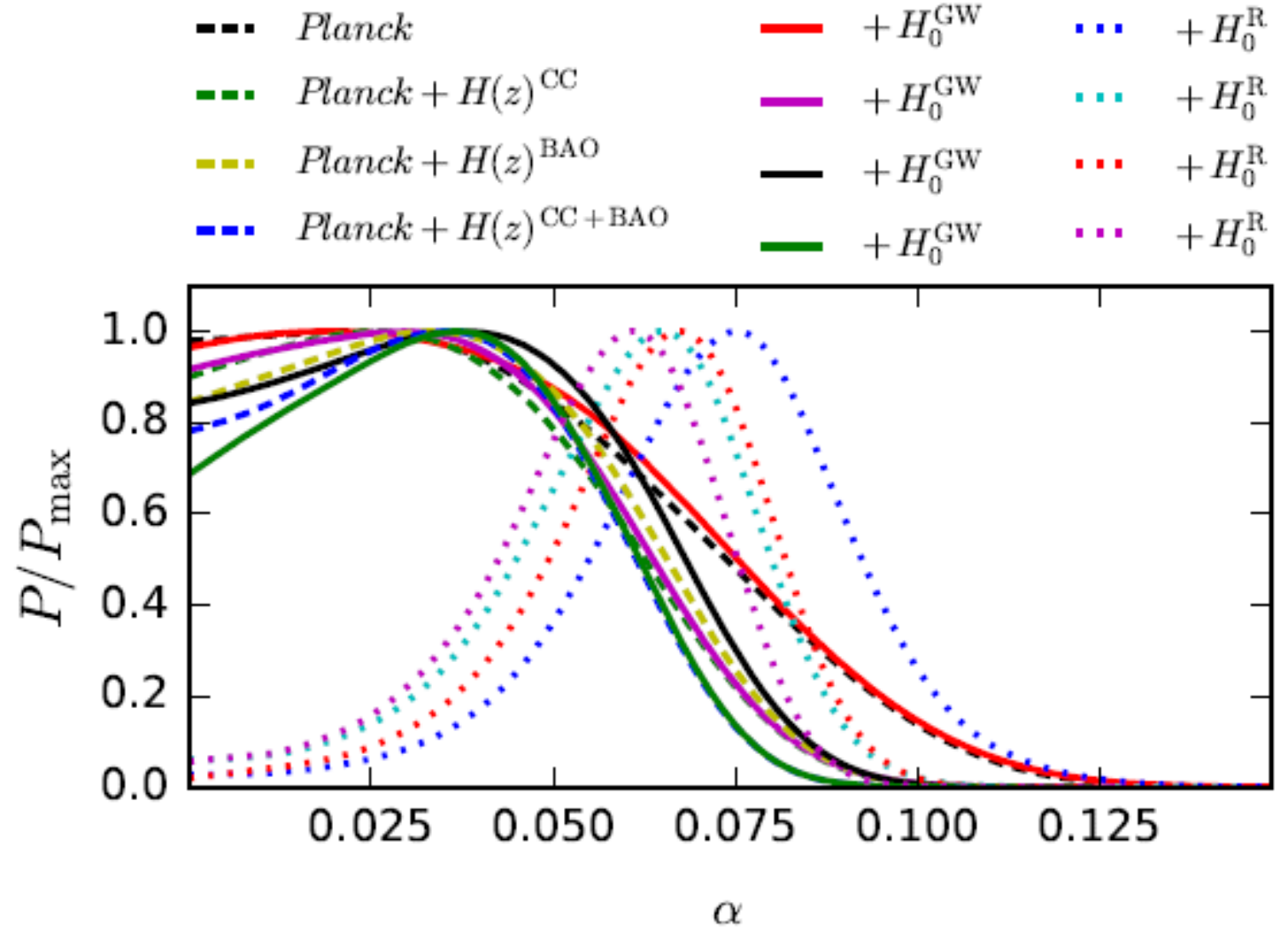}
\caption{Marginalised one--dimensional posterior distributions for the conformal coupling parameter $\alpha$, with the different data set combinations indicated in the figure. The respective parameter constraints are tabulated in Tables \ref{table:Tab1}--\ref{table:Tab3}.}  
\label{fig:1D_alpha}
\end{figure}

In order to assess the impact of independent measurements of the Hubble constant on the inferred model parameter constraints, we make use of a local measurement of the Hubble constant (hereafter denoted by $H_0^{\mathrm{R}}$) \citep{Riess:2016jrr} and the first gravitational--wave standard siren measurement (hereafter denoted by $H_0^{\mathrm{GW}}$) \citep{Abbott:2017xzu}. Since the latter marginalised posterior distribution for the Hubble constant is strongly non--Gaussian, we implemented this prior via an interpolating generalised normal distribution function that can adequately reproduce the reported constraint of \citet{Abbott:2017xzu}.

In addition, we occasionally further include information on the cosmic expansion history by making use of Hubble parameter measurements at several redshifts derived from the cosmic chronometers technique \citep{Simon:2004tf,Stern:2009ep,Moresco:2012jh,Zhang:2012mp,Moresco:2015cya,Moresco:2016mzx,Ratsimbazafy:2017vga} and also from BAO surveys \citep{2017A&A...603A..12B,2017A&A...608A.130D,Alam:2016hwk}, which we respectively refer to as $H(z)^{\mathrm{CC}}$ and $H(z)^{\mathrm{BAO}}$.

{\setlength\extrarowheight{5pt}
\setlength{\tabcolsep}{4.9pt}
\begin{table*}
\begin{center}
\caption{\label{table:Tab2} As in Table \ref{table:Tab1}, we here report the mean values and $1\sigma$ errors for each model parameter, together with the $1\sigma$ ($2\sigma$) upper limits of $\lambda$ and $\alpha$. The Hubble constant is given in units of km s$^{-1}$ Mpc$^{-1}$.}
\begin{tabular}{ p{1.85cm}  c  c  c  c  c  c } 
 \hline
\hline
Parameter~  &  ~\textit{Planck}$\,+\,H(z)^{\mathrm{CC}}$~ & ~$+\,H_0^{\mathrm{GW}}$~ & ~$+\,H_0^{\mathrm{R}}$~ & ~\textit{Planck}$\,+\,H(z)^{\mathrm{BAO}}$~ & ~$+\,H_0^{\mathrm{GW}}$~ & ~$+\,H_0^{\mathrm{R}}$~  \\[2.5pt]
\hline
100 $\Omega_b h^2$\dotfill & $2.2260^{+0.0164}_{-0.0166}$ & $2.2265^{+0.0159}_{-0.0166}$ & $2.2280^{+0.0172}_{-0.0171}$ & $2.2267^{+0.0163}_{-0.0167}$ & $2.2265^{+0.0166}_{-0.0167}$ & $2.2275^{+0.0168}_{-0.0172}$\\
$\Omega_c h^2$\dotfill & $0.11823^{+0.00226}_{-0.00169}$ & $0.11812^{+0.00226}_{-0.00170}$ & $0.11504^{+0.00206}_{-0.00203}$ & $0.11813^{+0.00231}_{-0.00165}$ & $0.11796^{+0.00237}_{-0.00167}$ & $0.11471^{+0.00205}_{-0.00199}$ \\
100 $\theta_s$\dotfill & $1.04180^{+0.00031}_{-0.00032}$ & $1.04180^{+0.00032}_{-0.00032}$ & $1.04190^{+0.00032}_{-0.00032}$ & $1.04180^{+0.00031}_{-0.00031}$ & $1.04180^{+0.00032}_{-0.00032}$ & $1.04180^{+0.00032}_{-0.00032}$ \\
$\tau_{\mathrm{reio}}$\dotfill & $0.062675^{+0.013844}_{-0.014080}$ & $0.062914^{+0.013913}_{-0.014234}$ & $0.065480^{+0.013984}_{-0.014079}$ & $0.063134^{+0.013886}_{-0.013882}$ & $0.063140^{+0.013985}_{-0.014053}$ & $0.064943^{+0.013900}_{-0.014071}$ \\
$\ln(10^{10}A_s)$\dotfill & $3.0580^{+0.0257}_{-0.0257}$ & $3.0585^{+0.0262}_{-0.0259}$ & $3.0632^{+0.0256}_{-0.0259}$ & $3.0590^{+0.0258}_{-0.0257}$ & $3.0589^{+0.0255}_{-0.0259}$ & $3.0624^{+0.0260}_{-0.0256}$ \\
$n_s$\dotfill & $0.96589^{+0.00486}_{-0.00506}$ & $0.96605^{+0.00485}_{-0.00511}$ & $0.96806^{+0.00477}_{-0.00490}$ & $0.96616^{+0.00486}_{-0.00495}$ & $0.96621^{+0.00487}_{-0.00497}$ & $0.96794^{+0.00474}_{-0.00489}$ \\ 
$\lambda$\dotfill & $0.76984^{+0.26304}_{-0.76978}$ & $0.72920^{+0.24056}_{-0.72919}$ & $0.41516^{+0.11982}_{-0.41516}$ & $0.84512^{+0.53738}_{-0.59177}$ & $0.78491^{+0.27099}_{-0.78490}$ & $0.41529^{+0.12331}_{-0.41528}$ \\ 
$\alpha$\dotfill & $0.033828^{+0.010437}_{-0.033828}$ & $0.034153^{+0.010661}_{-0.034152}$ & $0.058887^{+0.020880}_{-0.013856}$ & $0.035409^{+0.014387}_{-0.031933}$ & $0.036523^{+0.016976}_{-0.030794}$ & $0.062259^{+0.019390}_{-0.012494}$ \\
$\lambda$\dotfill & $<1.0329(1.5707)$ & $<0.9698(1.5203)$ & $<0.5350(0.9883)$ & $<1.3825(1.6113)$ & $<1.0559(1.5703)$ & $<0.5386(0.9587)$ \\ 
$\alpha$\dotfill & $<0.0443(0.0723)$ & $<0.0448(0.0724)$ & $<0.0798(0.0933)$ & $<0.0498(0.0733)$ & $<0.0535(0.0742)$ & $<0.0816(0.0950)$ \\[2pt]
\hline
$H_0$\dotfill & $67.450^{+2.5262}_{-2.0578}$ & $67.682^{+2.3382}_{-1.9280}$ & $70.994^{+1.5490}_{-1.5964}$ & $67.247^{+2.8785}_{-2.4777}$ & $67.611^{+2.6689}_{-2.3226}$ & $71.305^{+1.5829}_{-1.5888}$ \\
$\Omega_m$\dotfill & $0.31005^{+0.02010}_{-0.02741}$ & $0.30754^{+0.02038}_{-0.02481}$ & $0.27294^{+0.01535}_{-0.01637}$ & $0.31207^{+0.02484}_{-0.03151}$ & $0.30818^{+0.02304}_{-0.02911}$ & $0.26990^{+0.01489}_{-0.01642}$ \\
$\sigma_8^{}$\dotfill & $0.82419^{+0.02451}_{-0.02365}$ & $0.82632^{+0.02302}_{-0.02265}$ & $0.86038^{+0.02056}_{-0.02161}$ & $0.82253^{+0.02674}_{-0.02784}$ & $0.82597^{+0.02573}_{-0.02690}$ & $0.86413^{+0.02128}_{-0.02161}$ \\
$z_{\mathrm{reio}}$\dotfill & $8.4632^{+1.4174}_{-1.2493}$ & $8.4831^{+1.4188}_{-1.2634}$ & $8.6895^{+1.3937}_{-1.2363}$ & $8.5055^{+1.4117}_{-1.2340}$ & $8.5035^{+1.4198}_{-1.2442}$ & $8.6365^{+1.3874}_{-1.2283}$ \\[3.5pt]
\hline
\hline
\end{tabular}
\end{center}
\end{table*}}

We infer the parameter posterior distributions together with their confidence limits via a customised version of the Markov Chain Monte Carlo (MCMC) package \texttt{Monte Python} \citep{Audren:2012wb}, which is interfaced with a modified version of the cosmological Boltzmann code \texttt{CLASS} \citep{Blas:2011rf}, in which we evolve the background as well as the synchronous gauge linear perturbation equations \citep{Mifsud:2017fsy}. For our results, we also made use of the MCMC analysis package \texttt{GetDist} \citep{Lewis:2002ah}, and checked that the results are in an excellent agreement with those obtained from \texttt{Monte Python}.

\begin{figure}
\centering
  \includegraphics[width=0.99\columnwidth]{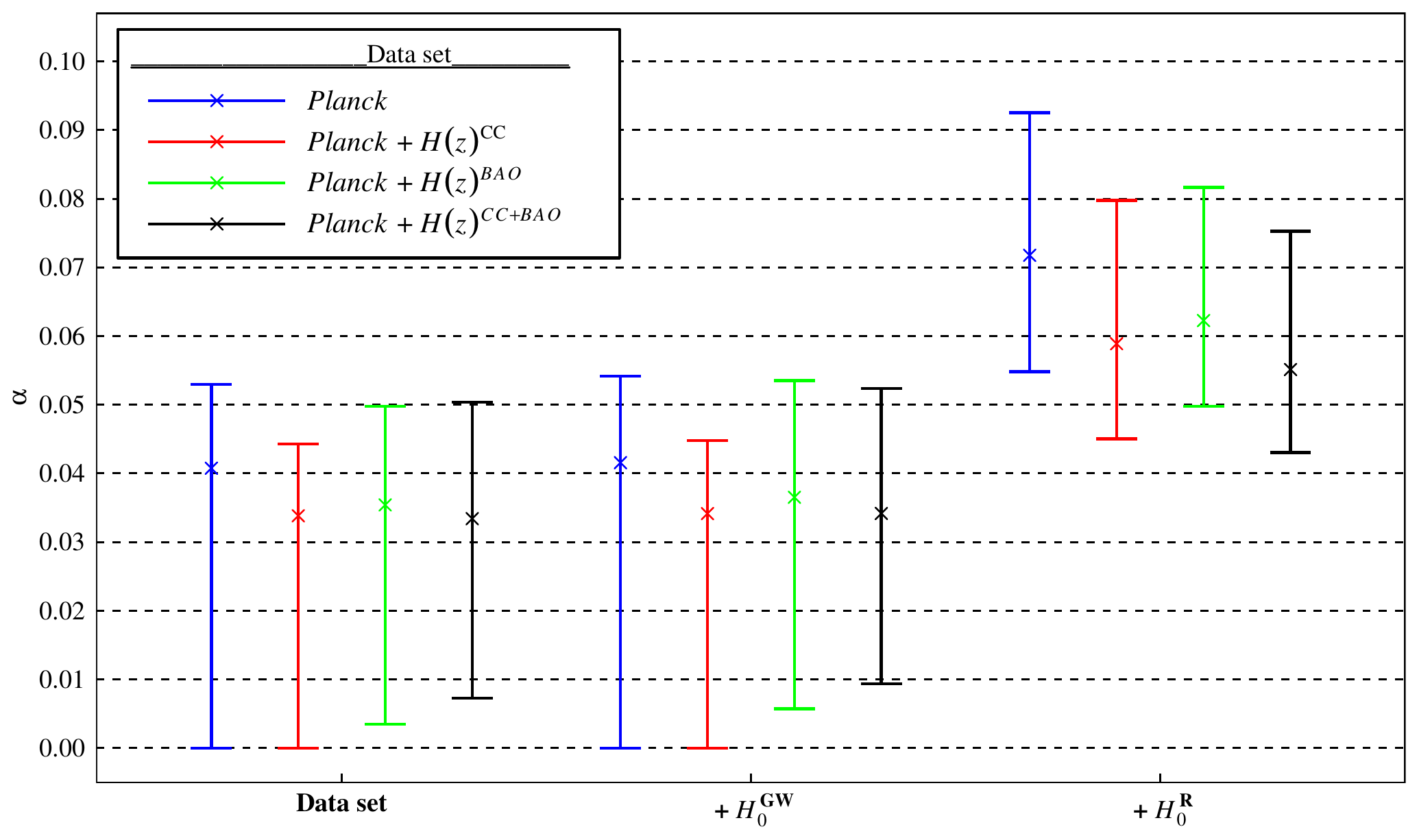}
\caption{The coloured intervals correspond to the inferred $1\sigma$ two--tail limits on the conformal coupling strength parameter $\alpha$. We illustrate all the data set combinations considered in this paper. }  
\label{fig:alpha_vs_H0_plot}
\end{figure}

We consider flat priors for the interacting DE model parameters that are allowed to vary in our MCMC analyses. The full range of each flat prior is listed in Table \ref{table:priors}. This set of parameters consists of $\Theta=\big\{\Omega_b h^2,\,\Omega_c h^2,\,100\,\theta_s,\,\tau_{\mathrm{reio}},\,\ln(10^{10}A_s),\,n_s,\,\lambda,\,\alpha\big\}$. Here, $h$ is defined in terms of the Hubble constant via $H_0=100\,h$ km s$^{-1}$ Mpc$^{-1}$, $\Omega_b h^2$ represents the effective fractional abundance of uncoupled baryons, $\Omega_c h^2$ is the pressureless coupled CDM effective energy density, $100\,\theta_s$ is the angular scale of the sound horizon at last scattering defined by the ratio of the sound horizon at decoupling to the angular diameter distance to the last scattering surface, $\tau_{\mathrm{reio}}$ is the reionization optical depth parameter, $\ln(10^{10}A_s)$ is the log power of the scalar amplitude of the primordial power spectrum together with its scalar spectral index $n_s$, $\lambda$ is the slope of the scalar field exponential potential, and $\alpha$ is the conformal coupling parameter. The inferred constraints on these parameters are reported in the top block of Tables \ref{table:Tab1}--\ref{table:Tab3}. Moreover, we also vary the nuisance parameters according to the procedure described in \citet{Ade:2015xua} and \citet{Aghanim:2015xee}. In the lower block of Tables \ref{table:Tab1}--\ref{table:Tab3} we present marginalised constraints on a number of derived cosmological parameters, including: $H_0$, the current total fractional abundance of non--relativistic matter $\Omega_m$, the linear theory rms fluctuation in total matter in $8\,h^{-1}$ Mpc spheres denoted by $\sigma_8^{}$, and the reionization redshift $z_\text{reio}$. We further adopt a pivot scale of $k_0=0.05\,\textrm{Mpc}^{-1}$, and we assume purely adiabatic scalar perturbations at very early times with null runnings of the scalar spectral index. Moreover, we fix the neutrino effective number to its standard value of $N_\mathrm{eff}=3.046$ \citep{Mangano:2001iu}, as well as the photon temperature today to $T_0=2.7255$ K \citep{Fixsen:2009ug}. As mentioned earlier, we assume spatial flatness.

{\setlength\extrarowheight{5pt}
\begin{table*}
\begin{center}
\caption{\label{table:Tab3} As in Tables \ref{table:Tab1} and \ref{table:Tab2}, we here report the mean values and $1\sigma$ errors for each model parameter, together with the $1\sigma$ ($2\sigma$) upper limits of $\lambda$ and $\alpha$. The Hubble constant is given in units of km s$^{-1}$ Mpc$^{-1}$.}
\begin{tabular}{ p{1.85cm} c  c  c } 
 \hline
\hline
Parameter~  &  ~\textit{Planck}$\,+\,H(z)^{\mathrm{CC+BAO}}$~ & ~$+\,H_0^{\mathrm{GW}}$~ & ~$+\,H_0^{\mathrm{R}}$~  \\[2.5pt]
\hline
100 $\Omega_b h^2$\dotfill & $2.2267^{+0.0162}_{-0.0168}$ & $2.2267^{+0.0162}_{-0.0167}$ & $2.2280^{+0.0167}_{-0.0174}$ \\
$\Omega_c h^2$\dotfill & $0.11830^{+0.00191}_{-0.00160}$ & $0.11819^{+0.00193}_{-0.00160}$ & $0.11554^{+0.00187}_{-0.00182}$ \\
100 $\theta_s$\dotfill & $1.04180^{+0.00032}_{-0.00032}$ & $1.04180^{+0.00032}_{-0.00031}$ & $1.04190^{+0.00031}_{-0.00032}$ \\
$\tau_{\mathrm{reio}}$\dotfill & $0.063140^{+0.013799}_{-0.014142}$ & $0.062965^{+0.013773}_{-0.014171}$ & $0.064670^{+0.013811}_{-0.014322}$ \\
$\ln(10^{10}A_s)$\dotfill & $3.0589^{+0.0254}_{-0.0259}$ & $3.0585^{+0.0258}_{-0.0257}$ & $3.0617^{+0.0259}_{-0.0258}$ \\
$n_s$\dotfill & $0.96599^{+0.00476}_{-0.00490}$ & $0.96601^{+0.00471}_{-0.00490}$ & $0.96752^{+0.00471}_{-0.00486}$ \\ 
$\lambda$\dotfill & $0.80183^{+0.31457}_{-0.76222}$ & $0.74427^{+0.24977}_{-0.74424}$ & $0.38575^{+0.11275}_{-0.38575}$ \\ 
$\alpha$\dotfill & $0.033412^{+0.016978}_{-0.026135}$ & $0.034171^{+0.018204}_{-0.024816}$ & $0.055134^{+0.020138}_{-0.012074}$ \\
$\lambda$\dotfill & $<1.1164(1.5774)$ & $<0.9940(1.5273)$ & $<0.4985(0.9075)$ \\ 
$\alpha$\dotfill & $<0.0504(0.0675)$ & $<0.0524(0.0680)$ & $<0.0753(0.0874)$ \\[2pt]
\hline
$H_0$\dotfill & $67.274^{+2.6365}_{-1.9748}$ & $67.582^{+2.4215}_{-1.7542}$ & $70.664^{+1.4042}_{-1.4232}$ \\
$\Omega_m$\dotfill & $0.31178^{+0.02019}_{-0.02794}$ & $0.30854^{+0.01886}_{-0.02474}$ & $0.27641^{+0.01407}_{-0.01482}$ \\
$\sigma_8^{}$\dotfill & $0.82242^{+0.02497}_{-0.02274}$ & $0.82519^{+0.02323}_{-0.02162}$ & $0.85648^{+0.01924}_{-0.02031}$ \\
$z_{\mathrm{reio}}$\dotfill & $8.5082^{+1.4049}_{-1.2510}$ & $8.4900^{+1.4034}_{-1.2569}$ & $8.6202^{+1.3869}_{-1.2623}$ \\[3.5pt]
\hline
\hline
\end{tabular}
\end{center}
\end{table*}}
%
%

\section{Results}
\label{sec:results}
We here discuss the inferred cosmological parameter constraints following the procedure described in Section \ref{sec:Data Sets}. As illustrated in the second column of Table \ref{table:Tab1}, the \textit{Planck} data set places tight limits on all model parameters, allowing only for a tiny conformal coupling within the dark cosmic sector. This is consistent with our previous analyses which we presented in \citet{vandeBruck:2017idm}, where we further showed that the inclusion of large--scale structure cosmic probes lead to tighter upper limits on $\alpha$, although a mild tension exists between some of these growth of structure data sets in the spatially--flat $\Lambda$CDM model \citep{Vikhlinin:2008ym,Henry:2008cg,Mantz:2009fw,Rozo:2009jj,Tinker:2011pv,Benson:2011uta,Hajian:2013rhm,Ade:2013lmv}. Thus, the cosmological bounds presented in this analysis should be considered as complementary and conservative. If we first focus on the Hubble constant constraints reported in Table \ref{table:Tab1}, we clearly observe that the \textit{Planck} data set prefers lower values of $H_0$ with respect to when we consider $H_0^{\mathrm{GW}}$, and particularly when we take into account the $H_0^{\mathrm{R}}$ prior. Consequently, a slightly larger dark sector coupling is allowed when we consider the \textit{Planck}$\,+\,H_0^{\mathrm{GW}}$ data set combination, although the inferred constraints are consistent with the \textit{Planck} data set constraints. However, the \textit{Planck}$\,+\,H_0^{\mathrm{R}}$ joint data set leads to a non--null dark sector coupling at a statistically high significance (see Fig. \ref{fig:1D_alpha} and Fig. \ref{fig:alpha_vs_H0_plot}). Due to the well--known correlation between the parameters $\alpha$ and $\sigma_8^{}$ \citep{vandeBruck:2017idm}, the \textit{Planck}$\,+\,H_0^{\mathrm{R}}$ bound on $\alpha$ gives rise to a significantly large value of $\sigma_8^{}$ which might not be fully--compatible with probes of the large--scale structure (see \citet{vandeBruck:2017idm} for a detailed discussion).

\begin{figure}
\centering
  \includegraphics[width=0.99\columnwidth]{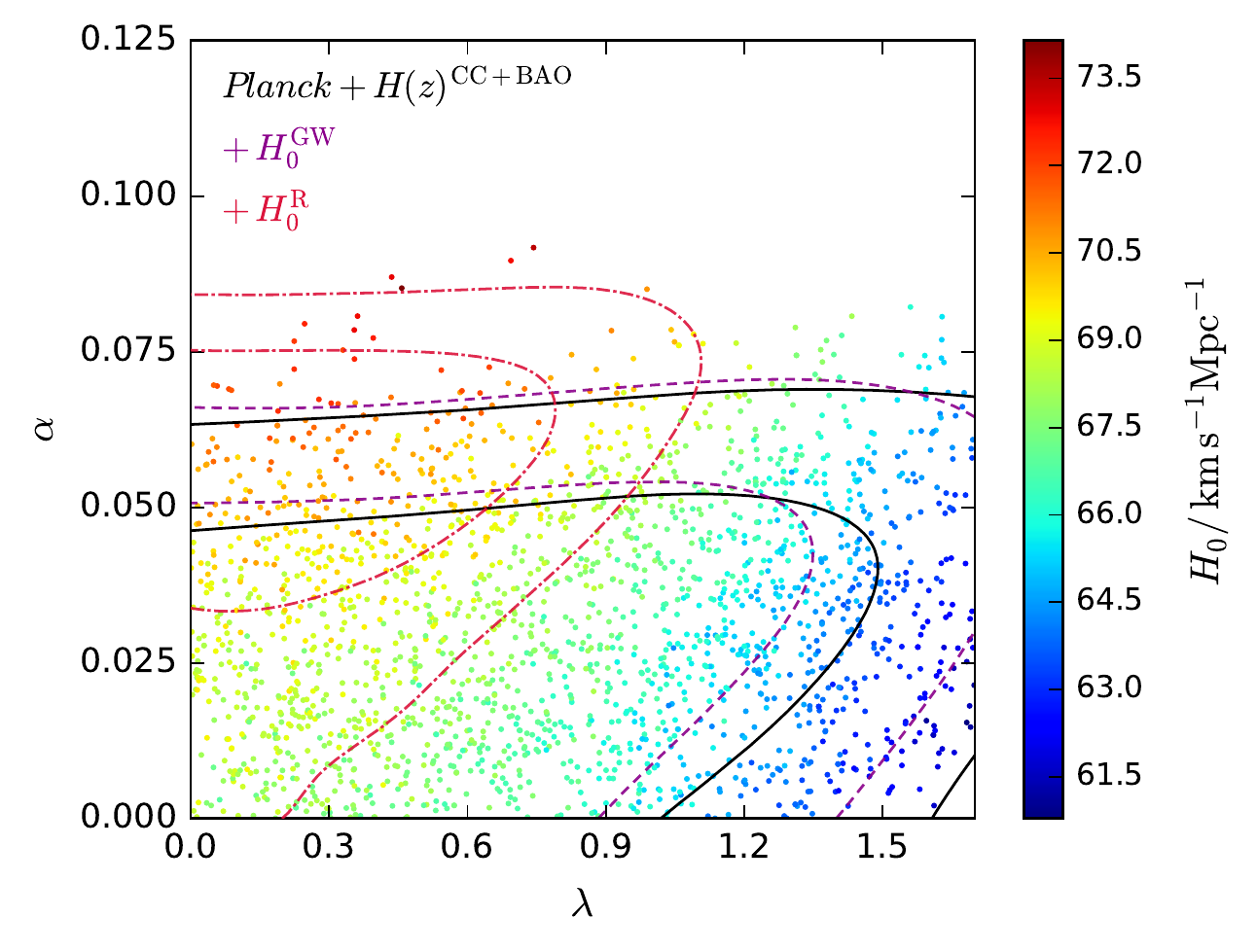}
\caption{Marginalised two--dimensional constraints on the parameters $\lambda$ and $\alpha$, together with samples from the \textit{Planck}$\,+\,H(z)^{\mathrm{CC+BAO}}$ joint data set colour coded with the value of the Hubble constant.}  
\label{fig:3D}
\end{figure}

Since the CMB anisotropies mainly probe the high--redshift Universe, we further add some information about the low--redshift cosmic expansion by considering the $H(z)^{\mathrm{CC}}$ and $H(z)^{\mathrm{BAO}}$ data sets, as described in Section \ref{sec:Data Sets}. We present these constraints in Table \ref{table:Tab2}, where we independently consider the $H(z)^{\mathrm{CC}}$ and $H(z)^{\mathrm{BAO}}$ data sets along with the \textit{Planck} data set and the Hubble constant priors, whereas in Table \ref{table:Tab3} we jointly consider the Hubble parameter measurements (hereafter denoted by $H(z)^{\mathrm{CC+BAO}}$). The consideration of these cosmic expansion measurements lead to improved constraints on the interacting DE model parameters with respect to the inferred constraints from the \textit{Planck} data set only. As we clearly illustrate in Figs. \ref{fig:1D_alpha} and \ref{fig:alpha_vs_H0_plot}, the $H_0^{\mathrm{GW}}$ prior always gives consistent constraints on $\alpha$ with those derived from the \textit{Planck}$\,+\,H(z)^{\mathrm{CC}}$, \textit{Planck}$\,+\,H(z)^{\mathrm{BAO}}$, and \textit{Planck}$\,+\,H(z)^{\mathrm{CC+BAO}}$ data set combinations. On the other hand, the $H_0^{\mathrm{R}}$ prior is always found to be associated with a non--null conformal coupling between DM and DE, although the inclusion of the cosmic expansion data sets lead to slightly smaller values of $\alpha$ with respect to the \textit{Planck} only constraints, but still not consistent with a vanishing dark sector coupling.

\begin{figure}
\centering
  \includegraphics[width=0.99\columnwidth]{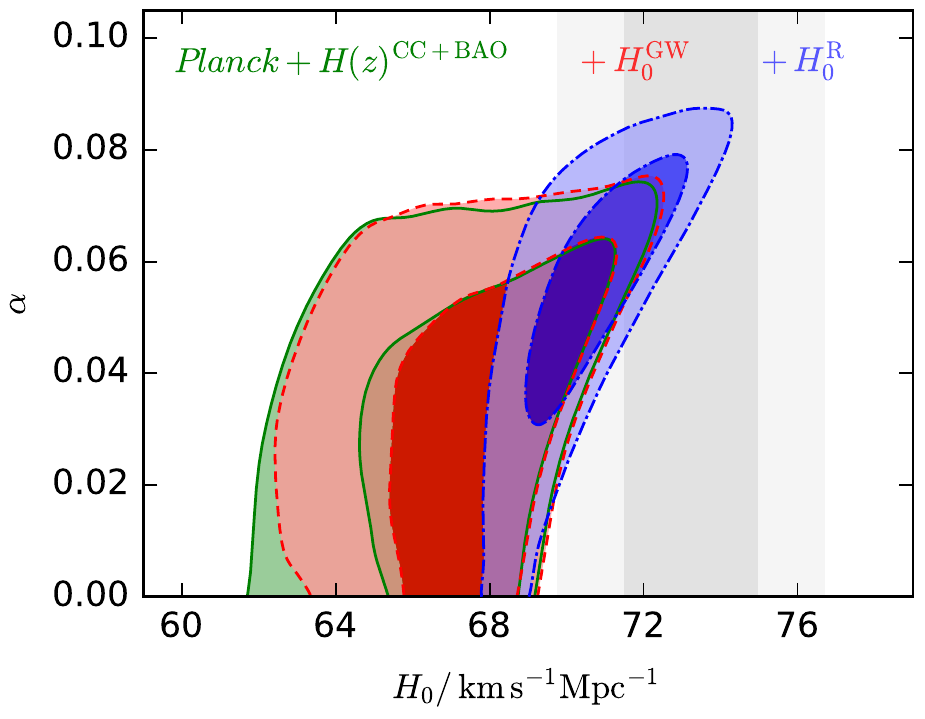}
\caption{Marginalised two--dimensional likelihood constraints on the Hubble constant and the conformal coupling parameter $\alpha$ with different data set combinations indicated in the figure. The gray (light gray) band shows the $1\sigma$ ($2\sigma$) constraint on the Hubble constant as reported in \citet{Riess:2016jrr}.}  
\label{fig:2D}
\end{figure}

Moreover, the $H_0$ likelihood priors improve the upper limit on the scalar field exponential potential parameter $\lambda$, particularly when we consider the $H_0^{\mathrm{R}}$ measurement in our data set combinations. This is depicted in Fig. \ref{fig:3D}, where we show the two--dimensional posteriors for the \textit{Planck}$\,+\,H(z)^{\mathrm{CC+BAO}}$, \textit{Planck}$\,+\,H(z)^{\mathrm{CC+BAO}}\,+\,H_0^{\mathrm{GW}}$, and \textit{Planck}$\,+\,H(z)^{\mathrm{CC+BAO}}\,+\,H_0^{\mathrm{R}}$ data sets in the $\lambda$--$\alpha$ plane, along with colour coded samples depicting the value of the Hubble constant. In Fig. \ref{fig:2D} we show the marginal correlation between $\alpha$ and $H_0$ (consistent with \citet{vandeBruck:2017idm}), where we present the two--dimensional likelihood constraints in the $H_0$--$\alpha$ plane. From Figs. \ref{fig:3D} and \ref{fig:2D}, we can clearly see the consistency between the inferred constraints in the $\lambda$--$\alpha$ and $H_0$--$\alpha$ planes with the \textit{Planck}$\,+\,H(z)^{\mathrm{CC+BAO}}$ and \textit{Planck}$\,+\,H(z)^{\mathrm{CC+BAO}}\,+\,H_0^{\mathrm{GW}}$ joint data sets. Since the cosmic distance ladder measurement of the Hubble constant is more accurate than the gravitational--wave standard siren measurement, such that the latter is compatible with a broad range of $H_0$ values, it is expected that tighter constraints on $H_0$ are derived in our interacting DE model when we make use of the $H_0^{\mathrm{R}}$ likelihood prior. This is depicted in Fig. \ref{fig:2D}, where we also observe the preference for a non--null interaction in the dark cosmic sector. Unequivocally, independent constraints on the Hubble constant would be able to shed light on the nature of DE and DM, and provide complementary constraints to the forthcoming cosmological surveys which are forecasted \citep{Amendola:2011ie,Casas:2015qpa,Miranda:2017rdk} to place very tight limits on this dark sector interaction.

%

\section{Conclusions}
\label{sec:conclusions}
We quantitatively examined the impact of independent Hubble constant measurements on a tightly constrained direct coupling between DM and DE. In our interacting DE model we specifically considered a conformal coupling within the dark cosmic sector, in which DE is described by a dynamical canonical scalar field and the gravitational attraction between the DM particles deviates from the standard one in General Relativity, such that the effective attraction is enhanced by the presence of a fifth--force. Since this coupling has been repeatedly shown to be robustly constrained by the \textit{Planck} CMB data set (see, for instance, \citet{Pettorino:2013oxa,Xia:2013nua,Ade:2015rim,Miranda:2017rdk,vandeBruck:2017idm}), we have always considered this crucial information in our joint data sets. Indeed, tight limits on all model parameters have been placed solely with the \textit{Planck} data set, including tight upper limits on the conformal coupling strength parameter $\alpha$, and the slope of the scalar field exponential potential $\lambda$. Moreover, we showed that the inclusion of a number of Hubble parameter measurements improve the \textit{Planck} only constraints.

In all our analyses which further considered the gravitational--wave standard siren measurement of the Hubble constant $H_0^{\mathrm{GW}}$, we found that this likelihood prior is compatible with a slightly larger conformal coupling within the dark cosmic sector, and marginally improves the upper limits on $\lambda$. Thus, we expect that near future standard siren measurements of the Hubble constant would place tighter constraints on this interacting DE model. Furthermore, this Hubble constant prior was not found to shift the model parameter constraints inferred from the more established cosmological data sets, and could therefore be considered as a conservative likelihood prior.

On the other hand, the inclusion of the more precise measurement of the Hubble constant derived from observations of Cepheid variables \citep{Riess:2016jrr} was always found to be characterised by a non--null interaction between DM and DE in the framework of the considered interacting DE model, as it can be seen from the blue region in Fig. \ref{fig:2D}. Although our results indicate that there is a strong preference for a non--vanishing dark sector coupling for this choice of data, it is important to notice that there are several precise cosmological data sets that can provide much tighter constraints on the model parameters and presumably not compatible with such a large coupling \citep{Miranda:2017rdk,vandeBruck:2017idm}. Thus, prospective data from CMB experiments and galaxy surveys, along with more observations of standard sirens which would be able to improve current estimates on the Hubble constant, will certainly enhance our understanding of the dark cosmic sector and potentially resolve the several tensions present between a number of cosmological probes. 

\section*{Note added:}
While this paper was being written up, a new constraint on the Hubble constant was presented in \citet{Soares-Santos:2019irc}, from another gravitational--wave source, the binary black--hole merger GW170814. The value stated, $H_0 = 75.2^{+39.5}_{-32.4}~$km s$^{-1}$ Mpc$^{-1}$, has larger error bars than the one used in our analyses. Because of this, adding this supplementary observation will not alter the results presented in the paper here. But clearly the future is bright for multi--messenger astronomy. 

\section*{Acknowledgements}

The work of CvdB is supported by the Lancaster-Manchester-Sheffield Consortium for Fundamental Physics under STFC Grant No. ST/L000520/1.




\bibliographystyle{mnras}
\bibliography{fullbib} 








\bsp	
\label{lastpage}
\end{document}